\documentclass[aps, prd,10pt,notitlepage,nofootinbib,superscriptaddress,showkeys,twocolumn]{revtex4-1}
\usepackage{amstext,amsmath,amssymb,amsfonts,bbm, amsthm}
\usepackage[latin1]{inputenc}
\usepackage{fancyhdr}
\usepackage{graphicx}
\usepackage{hyperref}
\usepackage{tocvsec2}

 \usepackage[usenames,dvipsnames]{pstricks}
 \usepackage{epsfig}
 \usepackage{pst-grad} % For gradients
 \usepackage{pst-plot} % For axes

\usepackage{color}

\def\beq{\begin{equation}}
\def\be{\begin{equation}}
\def\ee{\end{equation}}
\def\bes{\begin{eqnarray}}
\def\ees{\end{eqnarray}}

\def\f{\frac}

\def\pp{\partial}

\def\R{{\mathcal R}}

\begin{document}
\vspace{-2cm}
\begin{flushright}AEI-2013-169\end{flushright}

%%%%%%%%%%%%%%%%%%%%%%%%%%%%%%%%%%%%%%%%%%%%%%%%%%%

\title{\large \bf New perspectives on Hawking radiation}
\author{{Matteo Smerlak}}\email{smerlak@aei.mpg.de}
\affiliation{Max-Planck-Institut f\"ur Gravitationsphysik, Am M\"uhlenberg 1, D-14476 Golm, Germany}
\author{{Suprit Singh}}\email{suprit@iucaa.ernet.in}
\affiliation{Inter-University Centre for Astronomy and Astrophysics, Ganeshkhind, Pune 411 007, India}

\date{\small\today}

%%%%%%%%%%%%%%%%%%%%%%%%%%%%%%%%%%%%%
\begin{abstract}\noindent
We develop an adiabatic formalism to study the Hawking phenomenon from the perspective of Unruh-DeWitt detectors moving along non-stationary, non-asymptotic trajectories. When applied to \emph{geodesic} trajectories, this formalism yields the following results: $(i)$ though they have zero acceleration, the temperature measured by detectors on circular orbits is \emph{higher} than that measured by static detectors at the same distance from the hole, and diverges on the photon sphere, $(ii)$ in the near-horizon region, both outgoing \emph{and} incoming modes excite infalling detectors, and, for highly bound trajectories $(E\ll1)$, the latter actually \emph{dominate} the former. We confirm the apparent perception of high-temperature Hawking radiation by infalling observers with $E\ll1$ by showing that the energy flux measured by these observers diverges in the $E\rightarrow$ limit. We close by a discussion of the role played by spacetime curvature on the near-horizon Hawking radiation.
\end{abstract}
%%%%%%%%%%%%%%%%%%%%%%%%%%%%%%%%%%%%%%
\maketitle

\section{Introduction}

Hawking has famously predicted \cite{HAWKING1974,S1975} that the gravitational collapse of a (say spherically symmetric) matter distribution of mass $M$ will be perceived by observers at future infinity in the form of a stationary, outgoing thermal flux of massless particles with temperature $T_{H}=(8\pi M)^{-1}$. Like other general-relativistic effects, this ``Hawking phenomenon'' is \emph{a priori} observer-dependent, and it is not immediately clear what different observers, especially non-asymptotic ones, would measure in this context. To investigate this question, Unruh introduced in \cite{Unruh1976} a very enlightening approach, based on a simple particle detector model (referred to as an ``Unruh-Dewitt (UDW) detector''). In the case of static UDW detectors at ``radius'' $r$ in a Schwarzschild spacetime, a classic computation shows that the temperature measured by UDW detectors is $T_{H}(1-2M/r)^{-1/2}$(see e.g. \cite{Singleton:2011vj}); this shows that Hawking radiation satisfies the Tolman equilibrium condition \cite{Tolman1930}, just like a normal thermal bath.%, and confirms that Hawking radiation appears thermal not just at infinity, but in the whole outside region of the black hole.     

Another interesting class of trajectories for UDW detectors are \emph{infalling geodesics}. Since these are not tangent to a timelike Killing field, one cannot use any Tolman-like \emph{a priori} argument to infer the temperature measured by UDW detectors along these trajectories, and in particular when they cross the Schwarzschild horizon. Would they record Hawking radiation there? Or would they not, because Hawking radiation is created at some distance away from the black hole \cite{Unruh:1977uu}? 

%Answering this question with full clarity---at least within the usual semiclassical approximation---is especially timely, because 

Unruh addressed this question in \cite{Unruh1976}. Taking his cues from the geometric similarity between the Schwarzschild and Rindler horizons, and the observation that the Unruh temperature $T_{U}=a/2\pi$ of geodesic observers in Rindler spacetime is zero (because their acceleration $a$ is zero), he argued that ``a geodesic detector near the horizon will not see the Hawking flux of particles''. This conclusion---although not supported by an explicit calculation in \cite{Unruh1976}---fits with the general view put forward in that paper that, near the horizon, the in-vacuum of gravitational collapse is not different from the Minkowski vacuum. 

The reduction of the (curved spacetime) Hawking effect to the (flat spacetime) Unruh effect in the near-horizon region \cite{Jacobson} is further supported by the following fact: the Hawking temperature $T_{H}(1-2M/r)^{-1/2}$ perceived by a static observer at a radius $r>2M$ in Schwarzschild spacetime approaches the Unruh temperature $(M/r^{2})(1-2M/r)^{-1/2}/2\pi$ for a observer with the same acceleration in Minkowski spacetime when $R\rightarrow2M$. Singleton and Wilburn have described this fact by saying that ``the equivalence principle is restored at the horizon'' \cite{Singleton:2011vj}.

For all that, heuristic arguments as well as actual computations have recently challenged this conclusion. In the former category, Helfer has argued \cite{HELFER2004} that ``the vicinity of a black hole is a region in which essentially quantum-gravitational, Planck-scale, physics must dominate''; for Almheiri \emph{et al.} \cite{Almheiri2012}, the principle of information conservation and other quantum-information-theoretic constraints suggest the presence of a ``firewall'' at the horizon. While interesting in themselves, these arguments remain unfortunately too sketchy to lead to a definite prediction concerning the fate of horizon-crossing geodesic detectors. Following a more conventional semi-classical approach, Barbado \emph{et al.} \cite{Barbado2011} recently devised a framework, based on the notion of ``effective temperature'', to study the nature of Hawking radiation near the horizon more explicitly. Somewhat surprisingly, they found that the effective temperature of detector dropped with zero initial velocity from infinity will keep rising, reaching the value $4T_{H}$ at the horizon crossing. This result appears to contradict Unruh's aforementioned conclusion \cite{Unruh1976}, and does not accredit the notion that the Hawking effect reduces to the Unruh effect in the near-horizon region (at least in the sense that acceleration and temperature of UDW detectors are proportional).

%Although the authors of \cite{Barbado2011} give a physical argument supporting their finding, in terms of an infinite ``Doppler effect'' at the horizon, the perception of a finite temperature by near-horizon geodesic observers has not been considered firmly established by the community. In the author's experience, the consensus remains that Unruh's picture of the near-horizon physics is the correct one.

%Heuristic arguments as well as explicit computations have recently challenged this conclusion. In the former category, Helfer has argued \cite{HELFER2004} that ``the vicinity of a black hole is a region in which essentially quantum-gravitational, Planck-scale, physics must dominate''; for Almheiri \emph{et al.} \cite{Almheiri2012}, the principle of information conservation and other quantum-information-theoretic constraints suggest the presence of a ``firewall'' at the horizon. While interesting in themselves, these arguments are unfortunately too vague to lead to a definite prediction concerning the fate of horizon-crossing geodesic detectors. Following a more conventional semi-classical approach, 

Very appealing for its conceptual simplicity, the ``effective temperature'' approach of \cite{Barbado2011} has several technical limitations which could legitimate some skepticism about this result. First, the actual response function of a trans-horizon geodesic detector is \emph{not} explicitly evaluated in \cite{Barbado2011}, nor is any solid argument given to the effect that the ``effective temperature'' defined by these authors is indeed the one measured by an Unruh-DeWitt detector. Second, the framework used in \cite{Barbado2011}, being entirely based on the retarded Eddington-Finkelstein time coordinate $u$, does not apply to the interior region of the black hole; hence one could speculate that the effect found in \cite{Barbado2011} is perhaps an artifact of the horizon being singular with respect to the $u$-coordinate. Third, the vacuum state used in the definition of the ``effective temperature'' function in \cite{Barbado2011} is not the standard Unruh vacuum, but a new, unconventional state in the extended Schwarzschild geometry. One may fear that this new state is perhaps unsuitable to describe the physics of Hawking radiation. Fourth---and more crucially--- the fact that a radially infalling geodesic detector clicks at the horizon, and even records a finite temperature there, does not mean by itself that it perceives ``Hawking radiation''. The local spacetime geometry is not stationary along a radial geodesic, and actually changes more and more rapidly as the horizon (and then the singularity) is approached; it could be that the temperature measured at the horizon is due to this (trivial) curvature effect, and \emph{not} to the peculiar structure of the vacuum which is the true origin of Hawking radiation. Rather than ``does a near-horizon detector record a non-zero temperature?'', the question which should be addressed is therefore ``does a near-horizon geodesic detector record a non-zero temperature that cannot be explained away by a time-varying gravitational potential''? 

The purpose of this paper is to shed more light on these questions by reconsidering the response of geodesic UDW detectors during and after gravitational collapse. We address the four concerns above as follows:
\begin{enumerate}
\item
We define and study a suitably defined time-dependent ``temperature'' function, as in \cite{Barbado2011}, but also related it explicitly to UDW response functions by an \emph{adiabatic expansion}.
\item
We only use globally defined coordinates, regular on the horizon as well as everywhere else (except at the singularity).
\item
We avoid working with the extended Schwarzschild spacetime, where the effect of gravitational collapse must be mimicked by a suitable choice of vacuum state (the ``Unruh vacuum'' \cite{Unruh1976}). Instead, we consider gravitational collapse geometries and the corresponding---uniquely defined---in-vacuum state. 
\item
In the final section, we also consider generalized (non-Schwarzschild) black hole geometries, where the degeneracy between the surface gravity and the curvature scale at the horizon is lifted. This allows us to disentangle curvature effects and Hawking radiation at the horizon.  
\end{enumerate}

%In his ``Notes on black hole evaporation'' , Unruh elaborated on the analogy 

%Recently, however, this conclusion has been challenged by several authors, on various grounds. Studying a ``non-stationary vacuum'' mimicking gravitational collapse, Barbado \emph{et al.} \cite{Barbado2011} have found that ``generic freely-falling observers do not perceive vacuum when crossing the horizon, but an effective temperature a few times larger than the one that they perceived when it started to free-fall''. More radically, some authors has discussed the possible existence of a ``firewall'' \cite{Almheiri2012} or strong quantum-gravitational effects \cite{HELFER2004} in the neighborhood of the horizon. All in all, it is fair to say that the fate of an infalling observer as she crosses the horizon remains unknown. %Here we would like to suggest that the following observation might bring some clarit 

Another assumption made in \cite{Barbado2011} which we relax in this paper is that \emph{only outgoing modes couple to UDW detectors}. In fact, we will see that there exists a class of infalling trajectories for which the response of UDW detectors is actually \emph{dominated} by incoming modes. This is a somewhat surprising upshot of our analysis, which, to our knowledge, was not anticipated in the literature. 

%To illustrate this generality---and out of curiosity---we have also considered a simple class of non-radial geodesics, namely circular orbits. An interesting result which, to our knowledge, has not been previously reported in the literature, comes out of this analysis: geodesic observers on circular orbits around a Schwarzschild black hole measure a temperature \emph{higher} than static detectors at the same distance from the hole; this temperature diverges as the last circular orbit, aka the photon sphere ($R=3M$) is approached. This means that, as far as circular orbits are concerned, the photon sphere is a fiery place---indeed a firewall. (Of course, the unstability of Schwarzschild circular orbits for $R\leq6M$ limits the physical relevance of this divergence; we view it as a theoretical curiosity.)

The plan of the paper is as follows. In sec. \ref{genH}, we introduce the model of gravitational collapse used in the paper (the Vaidya ingoing shell), and discuss our ``quasi-temperature formalism'' to study the response of UDW detectors on arbitrary trajectories in that spacetime. We apply this formalism in sec. \ref{geo} to various geodesic trajectories: circular orbits, radially infalling trajectories, and inspiral orbits; our findings are tested against a flux computation in sec. \ref{flux}. In sec. \ref{curv}, we address in more detail the role played by the local spacetime curvature on the response of near-horizon geodesic detectors, by means of an artificial black hole model where the curvature vanishes near the horizon. Sec. \ref{disc} contains a discussion of our results and our conclusion. Details on adiabatic expansions of UDW-like response functions are given in Appendix B.  

\section{The Hawking effect along general trajectories}\label{genH}

The Hawking effect is the perception of an outgoing thermal flux at temperature $T_{H}\equiv(8\pi M)^{-1}$ by asymptotic inertial observers at rest relative to a Schwarzschild black hole. In this section we introduce a formalism to study response of particles detectors moving along more general trajectories. This allows us to consider two aspects of the Hawking phenomenon which are not often described in the literature: the \emph{onset} of black hole evaporation after the horizon has formed, and the \emph{time-dependence} of the spectra recorded by non-stationary detectors. 

Throughout this paper, we shall make the following assumptions and approximations: 
\begin{itemize}
\item
We only consider minimally coupled massless scalar fields.
\item
We neglect the contribution of non-spherically symmetric ($l\neq0$) field modes as well as all backscattering effects, so that the dynamics of the field is effectively two-dimensional.
\item
We neglect any backreaction of Hawking radiation on the background spacetime.
\end{itemize}
%All of these assumptions are classic in the context of ``semiclassical'' Hawking radiation.

\subsection{Collapse geometry and null coordinates}\label{collapse}

\vspace{-1cm}
\begin{center}
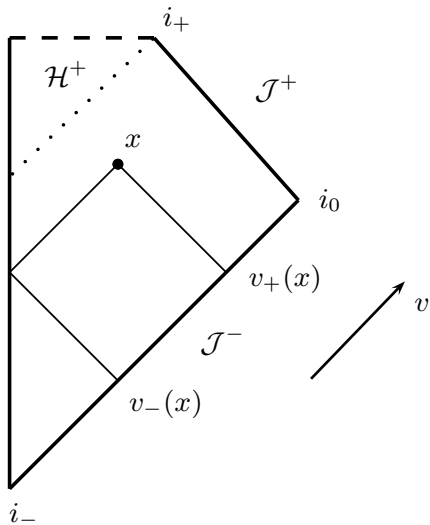
\begin{figure}[t!]
% Generated with LaTeXDraw 2.0.8
% Sun Dec 30 11:19:37 GMT+05:30 2012
% \usepackage[usenames,dvipsnames]{pstricks}
% \usepackage{epsfig}
% \usepackage{pst-grad} % For gradients
% \usepackage{pst-plot} % For axes
\scalebox{1.2} % Change this value to rescale the drawing.
{
\begin{pspicture}(0,-2.958047)(5.50291,2.958047)
\psline[linewidth=0.04cm](0.6210156,2.5596485)(0.6210156,-2.4403515)
\psline[linewidth=0.04cm](0.6210156,-2.4403515)(3.8210156,0.75964844)
\psline[linewidth=0.04cm](3.8210156,0.75964844)(2.2210157,2.5596485)
\psline[linewidth=0.04cm,linestyle=dashed,dash=0.16cm 0.16cm](0.6210156,2.5596485)(2.2210157,2.5596485)
\psdots[dotsize=0.12](1.8210156,1.1596484)
\psdots[dotsize=0.12](1.8210156,1.1596484)
\psline[linewidth=0.02cm](1.8210156,1.1596484)(3.0210156,-0.040351562)
\psline[linewidth=0.02cm](1.8210156,1.1596484)(0.6210156,-0.040351562)
\psline[linewidth=0.02cm](0.6210156,-0.040351562)(1.8210156,-1.2403516)
\usefont{T1}{ptm}{m}{n}
\rput(1.9924707,1.4246484){$x$}
\usefont{T1}{ptm}{m}{n}
\rput(3.6724708,-0.115351565){$v_{+}(x)$}
\usefont{T1}{ptm}{m}{n}
\rput(2.3524706,-1.4953516){$v_{-}(x)$}
\psline[linewidth=0.04cm,linestyle=dotted,dotsep=0.16cm](2.2210157,2.5596485)(0.64101565,1.0396484)
\usefont{T1}{ptm}{m}{n}
\rput(0.7824707,-2.7353516){$i_-$}
\usefont{T1}{ptm}{m}{n}
\rput(2.4824708,2.7646484){$i_+$}
\usefont{T1}{ptm}{m}{n}
\rput(4.1724706,0.76464844){$i_0$}
\usefont{T1}{ptm}{m}{n}
\rput(2.9824708,-0.79535156){$\mathcal{J}^{-}$}
\usefont{T1}{ptm}{m}{n}
\rput(3.5824707,2.0046484){$\mathcal{J}^{+}$}
\psline[linewidth=0.03cm,arrowsize=0.05291667cm 2.0,arrowlength=1.4,arrowinset=0.4]{->}(3.9610157,-1.2203516)(5.0010157,-0.14035156)
\usefont{T1}{ptm}{m}{n}
\rput(5.1924706,-0.39535156){$v$}
\usefont{T1}{ptm}{m}{n}
\rput(1.2524707,2.1646485){$\mathcal{H}^+$}
\end{pspicture} 
}
\caption{Definition of the (globally defined) ``eikonal coordinates'' $(v_{+},v_{-})$ for spherically-symmetric gravitational collapse.}
\label{penrosediag}
\end{figure}
\end{center}

Our model of gravitational collapse in this paper is the \emph{Vaidya ingoing shell}. (Another collapse geometry will be discussed in sec. \ref{curv}.) As is well known, this spacetime consists of two patches separated by a null thin-shell: a flat region inside the shell, and a Schwarzschild region outside the shell. Its metric is conveniently written in Eddington-Finkelstein advanced coordinates $(v,r)$ as 
\be\label{vaidya}
ds^{2}=-\left(1-\f{r_{s}}{r}\Theta(v)\right)dv^{2}+2dvdr+r^{2}d\Omega^{2},
\ee
where $r_{s}\equiv2M$ is twice the mass of the shell (with equation $v=0$), $\Theta(v)$ is the Heaviside function and $d\Omega^{2}=d\theta^{2}+\sin^{2}\theta d\varphi^{2}$ is the standard angular metric. The surface gravity of the black hole is $\kappa\equiv(2r_{s})^{-1}$. 

To analyze the Hawking phenomenon in such a collapse model, it useful to introduce globally defined \emph{null} coordinates $(v_{+},v_{-}$) for the $(v,r)$ sector of spacetime. These are constructed as follows. For each event $x$, consider the two null rays (incoming and outgoing) meeting at $x$, and define $v_{+}(x)$ and $v_{-}(x)$ as their respective Eddington-Finkelstein advanced time at past null infinity, with $v_{-}(x)\leq v_{+}(x)=v(x)$ (see Penrose diagram in Fig. \ref{penrosediag}). 

The physical interpretation of the $(v_{+},v_{-})$ coordinates is tied to the eikonal approximation for the propagation of massless fields $\phi(x)$, according to which
\be
\f{\phi(x)}{r}\sim\lim_{r\rightarrow\infty}\f{\phi(v_{+}(x),r)-\phi(v_{-}(x),r)}{r}.
\ee
In this approximation, the field at a given point $x$ is expressed as the superposition of two spherical waves emanating from past null infinity $\mathcal{J}^{-}$: a convergent wave, arriving directly from $v_{+}(v,r)$ (first term above), and a divergent wave, arriving from $v_{-}(v,r)$ after a reflection off the origin (second term above), as in Fig. \ref{penrosediag}. (For a harmonic mode $e^{-i\omega v}$ on $\mathcal{J}^{-}$, the phases of these two waves at a given point $x$ are $\omega v_{\pm}(x)$). For this reason, we call $(v_{+},v_{-})$ the \emph{eikonal coordinates} of $x$.

\begin{figure}[t!]
\centering
\includegraphics[scale=.95]{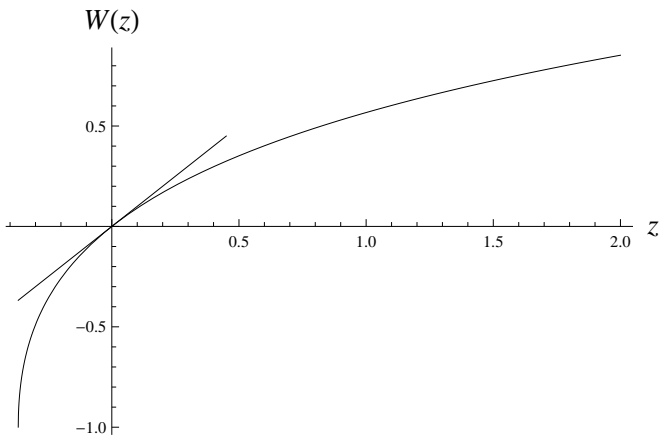}
\caption{The Lambert $W$-function, defined on $[e^{-1},\infty)$ by $W(z)e^{W(z)}=z$ and such that $W(z)\sim z$ for $z\rightarrow0$.}
\label{lambert}
\end{figure}

In the case of the Vaidya metric \eqref{vaidya}, it is possible to give the $(v,r)\mapsto(v_{+},v_{-})$ mapping in closed form. By construction, we have $v_{+}=v$, and, when $(v,r)$ is inside the shell, $v_{-}(v,r)=v-2r$; for a point $(v,r)$ outside the shell $(v>0)$, we must integrate the equation $ds^{2}=0$ from $(v,r)$ back to the point $(\widetilde{v}=0,\widetilde{r})$ where it meets the shell, and then write $v_{-}(v,r)=-2\widetilde{r}$. This gives
\begin{multline}\label{va}
v_{-}(v,r)=
\left
\{
\begin{array}{c}
v-2r\quad\quad\,\quad \qquad \quad \quad\quad\textrm{for}\ v<0\\
-2r_{s}\Big[1+W\left(\delta\,e^{\delta-\kappa v}\right)\Big]\quad\textrm{for}\ v\geq0,\end{array}
\right.
\end{multline}
where $\delta\equiv r/r_{s}-1$, $\kappa = 1/2r_s$ and $W(z)$  denotes the Lambert $W$-function, defined as the principal solution of $W(z)e^{W(z)}=z$. (For the reader's convenience, the graph of $W(z)$ plotted in Fig. \ref{lambert}.) 

\begin{figure}[t!]
\hspace{-.8cm}
\includegraphics[scale=.9]{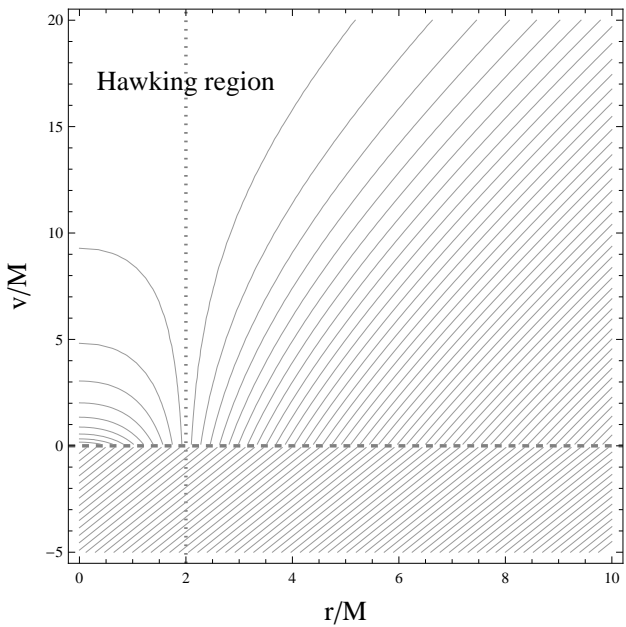}
\caption{``Portrait of the vacuum'': level curves of the eikonal coordinate $v_{-}(v,r)$ in the Vaidya spacetime. The dashed horizontal line is the collapsing shell, the dotted vertical line is the event horizon, and the region in the top-left corner is the ``Hawking region''.}
\label{fountain}
\end{figure}

Since $W(z)\sim z$ as $z\rightarrow0$, the relation \eqref{va} immediately gives the equation of the event horizon as $v_{-}=-2r_{s}$. More importantly, this relation also reveals the peculiar disruption of phase fronts induced by the gravitational collapse: from \eqref{va} we see that $v_{-}$ is constant where $\delta e^{\delta-\kappa v}$ is constant, and in particular $v_{-}\simeq -2r_{s}$ where $\delta e^{\delta-\kappa v}\ll1$. Due to the peculiar analytic form of the function $\delta\mapsto\delta e^{\delta-\kappa v}$ for each $v\geq0$, the region where $\delta e^{\delta-\kappa v}\ll1$ is very sharply defined. Inside this region, which expands away from the horizon as $v$ grows (until it eventually covers the whole of space), the phase of a vacuum fluctuation emerging from the shell after bouncing off its center is \emph{almost exactly equal} to $-2r_{s}$; this is where the Hawking phenomenon takes place. We call it the \emph{Hawking region}. The level curves of $v_{-}(v,r)$ are represented in Fig. \ref{fountain}; in pictorial terms, Fig. \ref{fountain} shows how gravitational collapse ``opens up'' the in-vacuum.

Notice that, in the Hawking region, the eikonal coordinate $v_{-}$ takes the same functional form as the Kruskal-Szekeres $U$ coordinate for eternal black holes. Defining indeed, \`a la Kruskal-Szekeres, $U(v,r)\equiv-2r_{s}e^{-\kappa u(v,r)}$ with $u(v,r)\equiv v-2r-2r_{s}\log\delta$, we read from see from \eqref{va} and the asymptotic formula $W(z)\sim z$ as $z\rightarrow0$ that
\be\label{late}
v_{-}(v,r)\sim U(v,r)-2r_{s}
\ee
when $\delta e^{\delta-\kappa v}\ll1$. The relation \eqref{late} can be thought of as providing the physical interpretation of Kruskal's and Szekeres' $U$ coordinate: $U$ captures the structure of outgoing phase fronts in the Hawking region of a (non-eternal) black hole.

%Relevant information about this function are given in Appendix. 

%Using the asymptotic formula $W(z)\sim z$ for $z\rightarrow0$ for the Lambert $W$-function, we can see that $v_{-}$ plays the same role for gravitational collapse as the Kruskal-Szekeres $U$ coordinate for eternal black holes. Indeed, if we define \`a la Kruskal-Szekeres $U(v,r)\equiv-2r_{s}e^{-\kappa u(v,r)}$ with $u(v,r)\equiv v-2r-2r_{s}\log\delta$, we see immediately that
%\be\label{late}
%v_{-}(v,r)\sim U(v,r)-2r_{s}
%\ee 
%near the horizon $(\vert\delta\vert\ll1)$ or in the asymptotic future $(\kappa v\gg\vert\delta\vert)$. In fact, one can show that, in these limits, \eqref{late} is the general form of the eikonal coordinate $v_{-}$ for a spherically symmetric black hole with surface gravity $\kappa$, whether Schwarzschild or not. We will see in sec. \ref{curv} that this universal behavior does \emph{not} mean that the response of Unruh-DeWitt detectors is the same on any such horizon, irrespective of the Riemann curvature there.  

\subsection{Unruh-DeWitt detectors}

Unruh-DeWitt (UDW) detectors \cite{Unruh1976,DeWitt1979} are point-like monopole detectors, which measure the Wightman function $G(x,y)$ of the field along a given trajectory $\gamma(\tau)$. To first order in perturbation theory, the response function of a UDW detector reads

\begin{widetext}
\be\label{response}
\mathcal{R}(\Omega)=2\,\textrm{Re}\,\int_{-\infty}^{\infty}du\,\chi(u)\int_{0}^{\infty}ds\, \chi(u-s)\,e^{-i\Omega s}\,G\big(\gamma(u),\gamma(u-s)\big).
\ee
\end{widetext}
Here, $\Omega$ is the energy gap between two stationary states of the detector, $G\big(\gamma(u),\gamma(u-s)\big)$ is the pull-back of the Wightman function to the detector's worldline, and $\chi(u)$ is a non-negative ``switching'' or ``window'' test function. The introduction of a switching function in the definition of the UDW response function ensures that $\mathcal{R}(\Omega)$ is well-defined and consistent with causality: choosing a switching function such that $\chi(u)\simeq 0$ for $u\geq\tau$ and $u\leq \tau-\Delta\tau$ ensures that $\mathcal{R}(\Omega)$ only depends on the time interval $\Delta\tau$ in the past of $\tau$. Motivated by this causality issue, Svaiter and Svaiter introduced in \cite{Svaiter1992} the simple ansatz $\chi(u)=\Theta(u-\tau-\Delta\tau)\Theta(\tau-u)$, which in the $\Delta\tau\rightarrow\infty$ limit leads to a transition rate ($\tau$-derivative of the response function $\R$)
\be\label{response1}
\dot{\R}(\Omega)=2\,\textrm{Re}\,\int_{0}^{\infty}ds\, e^{-i\Omega s} G\big(\gamma(\tau),\gamma(\tau-s)\big).
\ee
However appealing at first sight, this ansatz has two unpleasant features, which are due to the fact that $\chi(u)=\Theta(u-\tau-\Delta\tau)\Theta(\tau-u)$ is not a proper test function (it is not smooth): first, the standard $-i0$ prescription for the Wightman function yields non-Lorentz-invariant transition rates \cite{Schlicht2004} and must be replaced by a more complicated prescription \cite{Takagi1986,Schlicht2004,Langlois2006,Louko2008}; second, the ultraviolet behavior (large $\Omega$ limit) is qualitatively different on stationary and non-stationary trajectory \cite{Obadia:2007ds,Satz:2007uy}. Both these shortcomings disappear if a smooth switching function is used instead \cite{Higuchi:1993,Satz:2007uy}.

\subsection{Wightman function in the ($s$-wave) in-vacuum}

As mentioned earlier, in this paper we restrict our attention on the $s$-wave sector (spherically symmetric field configurations) of the in-vacuum. In this approximation, standard arguments \cite{Birrell1982,Helfer2003} show that $G(x,y)$ takes the logarithmic form
\be\label{2d}
G(x,y)\propto\ln\Big(\big(v_{+}(x)-v_{+}(y)-i0\big)\big(v_{-}(x)-v_{-}(y)-i0\big)\Big).
\ee
Using the relation \eqref{late}, this expression can be identified in the Hawking region with the one given in \cite{Unruh1976} to describe a radiating eternal black hole and known as the ``Unruh vacuum'',
\be\label{unruhvac}
G_{U}(x,y)\propto\ln\Big(\big(v(x)-v(y)-i0\big)\big(U(x)-U(y)-i0\big)\Big).
\ee

These states---the collapse vacuum \eqref{2d} and the Unruh vacuum \eqref{unruhvac}, in the $s$-wave sector---have a peculiar property, which is not shared by more general vacua: the incoming and outgoing modes couple to UDW detectors independently. This means that the UDW response function $\mathcal{R}$ splits as $\R=\R_{+}+\R_{-}$, where $\R_{\pm}$ is given by \eqref{response} with $G(x,y)$ replaced by
\be\label{log}
\ln\left(v_{\pm}\big(\gamma(u)\big)-v_{\pm}\big(\gamma(u-s)\big)-i0\right).
\ee
The thermal properties of $\R_{+}$ and $\R_{-}$ can thus be studied independently, and are completely determined by the behavior of each eikonal coordinate $v_{\pm}$ along the detector's trajectory. In particular, in the case of stationary trajectories leading leading to a thermal spectrum, the thermality (aka detailed balance) condition for $R_{+}$ and $\R_{-}$ defines \emph{two} temperatures $T_{+}$ and $T_{-}$, by
\be\label{DBC}
\R_{\pm}(-\Omega)=e^{\Omega/T_{\pm}}\R_{\pm}(\Omega).
\ee
These temperatures need not coincide in general. This is clear in the asymptotic Hawking effect, where $T_{+}=0$ and $T_{-}=T_{H}=\kappa/2\pi$.

\subsection{Hawking temperature from the peeling of outgoing modes}

The mechanism responsible for the perception of Hawking radiation has been described by many authors as an ``exponential redshift'' or ``peeling'' effect. After \cite{S1975}, this statement is usually expressed in terms of the canonical mapping from $\mathcal{J}^{+}$ to $\mathcal{J}^{-}$ which relates the retarded and advanced Eddington-Finkelstein times of a given null ray. From the perspective of UDW detectors, however, it is more natural to state the ``peeling'' condition in a trajectory-dependent way, as follows: if $\gamma(\tau)$ is the trajectory of an asymptotic observer at rest relative to the hole, the quantity $v_{-}(\gamma(\tau))$ satisfies
\be
-\f{\ddot{v}_-(\gamma(\tau))}{\dot{v}_-(\gamma(\tau))}=\kappa.
\ee
(From now on, we shall drop the explicit reference to $\gamma$, the trajectory being understood from the context.). Integrating this condition gives 
\be
\label{vminusspectrum}
v_{-}(\tau)-v_{-}(\tau-s)\propto e^{-\kappa(2\tau-s)/2}\sinh(\kappa s/2).
\ee
If we plug this relation into \eqref{log} and perform the standard residue integration  \cite{Birrell1982} (see Appendix~\ref{thermalresult} for details), we obtain the thermal spectrum at temperature $T_{H}=\kappa/2\pi$
\be\label{thermalspectrum}
\dot{\R}_{-}(\Omega)\propto\f{1}{\Omega(e^{2\pi\Omega/\kappa}-1)}. 
\ee
(The $\Omega\rightarrow0$ divergence is of course an artifact of the two-dimensional approximation.)

The main advantage of this trajectory-based formulation of the exponential redshift argument is that it applies \emph{mutatis mutandis} to any trajectory where $\ddot{v}_{-}/\dot{v}_{-}$ is a constant: for the same reasons that asymptotic observers such that $\vert\ddot{v}/\dot{v}\vert=\kappa$ perceive a thermal spectrum at temperature $\kappa/2\pi$, non-asymptotic observers such that $\vert\ddot{v}/\dot{v}\vert$ is constant perceive a thermal spectrum at temperature 
\be
T_{-}=\f{1}{2\pi}\Big|\f{\ddot{v}_{-}}{\dot{v}_{-}}\Big|.
\ee
This is the case e.g. for (late-time) static observers at any radius $r>r_{s}$: from the estimate \eqref{late}, we compute
\be
\Big|\f{\ddot{v}_{-}}{\dot{v}_{-}}\Big|=\kappa\dot{v}(r).
\ee
The time-derivative $\dot{v}(r)$ along a static trajectory can be read off from the metric \eqref{vaidya}, as $\dot{v}(r)=(1-r_{s}/r)^{-1/2}$, and therefore the temperature of outgoing modes measured by static UDW detector is readily found to be
\be\label{static}
T_{-}^{\textrm{stat}}(R)=T_{H}(1-r_{s}/r)^{-1/2}.
\ee
Again, this is consistent with the Tolman equilibrium condition \cite{Tolman1930}. 

\subsection{Non-static trajectories: quasi-temperature and adiabatic approximation}\label{quasi}

It has been argued in \cite{Barbado2011} that the \emph{approximate} constancy of $T_{-}(\tau)\equiv\vert\ddot{v}_{-}(\tau)/\dot{v}_{-}(\tau)\vert/(2\pi)$, expressed as 
\be
\eta_{-}\equiv\Big|\f{\dot{T}_{-}}{T_{-}^{2}}\Big|\ll1,
\ee
is a sufficiently condition for the perception of thermal Hawking radiation. This ``adiabaticity condition'', considered previously \cite{Barcelo2011a} in the context of evolving black holes, allows to study the response of UDW detectors along non-stationary trajectories in a very straightforward way, simply by computing $T_{-}(\tau)$. Heuristically, when $\eta_{-}\ll1$, the response function $\R(\tau,\Omega)$ of a UDW detector switched off (smoothly) at time $\tau$ will be indistinguishable from a thermal spectrum with $\tau$-dependent temperature $T_{-}(\tau)$. In fact, even when $\eta$ is not small, this relationship still holds in the ultraviolet limit, where $\Omega\gg T_{-}(\tau)$.

Following this rationale, we call \emph{quasi-temperature} of a time-dependent spectrum $\R(\tau,\Omega)$ a function $T(\tau)$ such that, at any given time $\tau$, the detailed balance condition holds in the ultraviolet limit,
\be
\R(\tau,-\Omega)\sim e^{\Omega/T(\tau)}\R(\tau,\Omega) \quad \textrm{for}\quad\vert\Omega\vert\gg T_{-}(\tau).
\ee
See Appendix B for more details on the adiabatic approximation. 

\subsection{Two quasi-temperatures for Hawking radiation}

Our approach to the study of Hawking radiation along general trajectories is based on this adiabatic approximation. For a given trajectory, we will compute the quasi-temperatures of both outgoing and incoming modes as
\be\label{quasiT}
T_{\pm}(\tau)=\f{1}{2\pi}\Big|\f{\ddot{v}_{\pm}(\tau)}{\dot{v}_{\pm}(\tau)}\Big|.
\ee
As we will see, these quasi-temperatures together with the corresponding adiabaticity parameters
\be
\eta_{-}\equiv\Big|\f{\dot{T}_{-}}{T_{-}^{2}}\Big|\ll1,
\ee
provide a valuable handle on the response of UDW detectors to Hawking radiation, especially in the large $\Omega$ limit.

%In the adiabatic limit where $\eta_{+}$ and $\eta_{-}$ are much smaller than $1$, the detailed balance condition not only in the ultraviolet limit but also around the characteristic frequency $\vert\Omega\vert\simeq T_{\pm}(\tau)$, and we may think of the quasi-temperatures as ``instantaneous temperatures''. 

\section{Response of geodesic detectors}\label{geo}

In this section, we apply the quasi-temperature formalism to the case of geodesic trajectories in the Schwarzschild region ($v>0$) of the Vaidya collapse geometry described above.

\begin{figure}[t!]
\begin{center}
\includegraphics[scale=1.1]{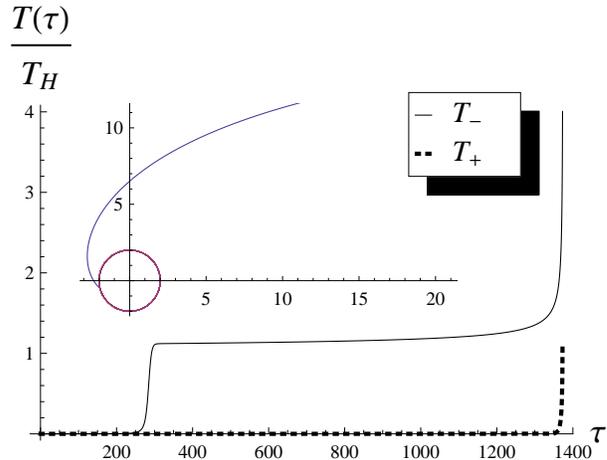}
\caption{The quasi-temperature of outgoing (continuous) and incoming (dashed) as a function of proper time along the inspiral geodesic $(E,L)=(1,3)$ until it reaches the horizon. The inset shows the geodesic in the equatorial plane. The step at $\tau\simeq300$ corresponds to the moment when the trajectory enters the Hawking region.}
\label{insp}
\end{center}
\end{figure}

\subsection{Schwarzschild geodesics}

Timelike Schwarzschild geodesics are characterized by two orbital parameters: their energy per unit rest mass $E$ and their angular momentum per unit rest mass $L=r^{2}\dot{\varphi}$. In Eddington-Finkelstein coordinates, the geodesic equations read
\be\label{system}
\left
\{
\begin{array}{c}
\dot{r}^{2}+(1-r_{s}/r)(1+L^{2}/r^{2})=E^{2}
\\
-(1-r_{s}/r)\dot{v}^{2}+2\dot{v}\dot{r}+L^{2}/r^{2}=-1.
\end{array}
\right.
\ee
From these equations and \eqref{va}, we can in principle obtain $v_{+}=v$, $v_{-}$ and their derivatives---hence the quasi-temperatures $T_{\pm}$ and the adiabaticity parameters $\eta_{\pm}$---along any geodesic $(E,L)$. This approach is illustrated in Fig. \ref{insp} for one inspiral trajectory with $(E,L)=(1,3)$, obtained by a numerical integration of \eqref{system}. This plot shows interesting phenomena: the setting off of Hawking radiation after gravitational collapse, the increase of the quasi-temperature of outgoing modes, \emph{and} of the quasi-temperature of incoming modes as the horizon is approached.

%
%
% $\ddot{v}_{+}=\ddot{v}$, $\dot{v}_{-}$ and $\ddot{v}_{-}$ (hence the quasi-temperatures $T_{\pm}$ and the corresponding adiabaticity parameters $\eta_{\pm}$) along any geodesic $(E,L)$ as functions of proper time. This approach is illustrated in fig. \ref{insp} for one inspiral trajectory, by means of a numerical integration of \eqref{system}. This plot shows interesting phenomena: the setting off of Hawking radiation after gravitational collapse, the increase of the quasi-temperature of outgoing modes \emph{and} of that of incoming modes as the horizon is approached, and their divergence on the singularity. 
%

In order to get an \emph{analytical} understanding of this plot, and in particular of the dependence of $T_{\pm}$ and $\eta_{\pm}$ on $E$ and $L$, we now consider special regimes of interest (circular orbits, radial trajectories in the asymptotic and near-horizon regimes) separately. Our presentation is  pedestrian. We have found, indeed, that the details of the computation of $T_{\pm}$ and $\eta_{\pm}$ provide a good deal of insight into the inner workings of the Hawking phenomenon: they display in a very transparent way the interplay between the geometry of spacetime (and its geodesics) and the structure of the in-vacuum which is responsible for the evaporation of black holes. %The results (in the late time limit, long after the shell has collapsed) are summarized below.

%In the following, with an emphasis on the relevant computation steps. We have found, indeed, that the computation of $T_{\pm}$ provides a good deal of insight into the inner workings of the Hawking phenomenon, with its interplay between the geometry of spacetime (and its geodesics) and the structure of the in-vacuum.

\subsection{Circular orbits}

\begin{figure}[t!]
\centering
\includegraphics[scale=1.1]{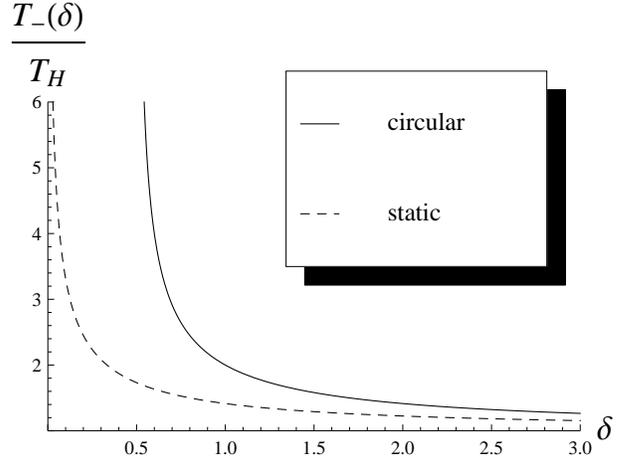}
\caption{Stationary trajectories: the temperature measured by static detectors (dashed line) and by geodesic detectors on a circular orbit (continuous line) around a Schwarzschild black hole as a function of $\delta=r/r_{s}-1$.}%The latter is always larger than the former and diverges on the photon sphere $(\delta=1/2)$.}
\label{circfig}
\end{figure}

Circular orbits are characterized by their Schwarzschild radius $r>3r_{s}/2$, or equivalently by $\delta=r/r_{s}-1>1/2$.\footnote{The corresponding energies and angular momenta are given by
$$
E_{\textrm{circ}}(r)=r_{s}\left[\f{\delta^{2}}{(\delta+1)(\delta-1/2)}\right]^{1/2},\ 
L_{\textrm{circ}}(r)=r_{s}\left[\f{(1+\delta)^{2}}{2(\delta-1/2)}\right]^{1/2}.
$$}
By virtue of their stationarity, these orbits have constant $\dot{v}$, given by 
\be\label{vdotcirc}
\dot{v}(\delta)=\left(\f{1+\delta}{\delta-1/2}\right)^{1/2}.
\ee
In particular, $\ddot{v}=0$, hence $T_{+}=0$: incoming modes do not couple to UDW detectors on circular orbits. On the other hand, from \eqref{va} we get for the outgoing modes
\be
\f{\ddot{v}_{-}(\delta)}{\dot{v}_{-}(\delta)}=-\kappa\dot{v}(\delta)\left(1+\delta e^{\delta-\kappa v}\f{W''(\delta e^{\delta-\kappa v})}{W'(\delta e^{\delta-\kappa v})}\right)
\ee
where primes denote derivative with respect to $z = \delta e^{\delta-\kappa v}$. Using \eqref{vdotcirc} and the standard formula for the derivative of the Lambert $W$-function, 
\begin{eqnarray}
W'(z)=\f{W(z)}{z\big(1+W(z)\big)},
\end{eqnarray}
we arrive at 
\begin{multline}\label{formulacirc}
T_{-}(v,\delta)=T_{H}\left(\f{1+\delta}{\delta-1/2}\right)^{1/2}
\times\\\left|1-\f{W(\delta e^{\delta-\kappa v})\big(2+W(\delta e^{\delta-\kappa v})\big)}{\big(1+W(\delta e^{\delta-\kappa v})\big)^{2}}\right|.
\end{multline}
In this expression, the second term represents the transient regime before the orbit has been ``swallowed'' by the Hawking region. After this transient, $T_{-}(v,\delta)$ reaches the stationary value 
\be
T^{\textrm{circ}}_{-}(\delta)\equiv\lim_{v\rightarrow\infty}T_{-}(v,\delta)=T_{H}\left(\f{1+\delta}{\delta-1/2}\right)^{1/2}.
\ee
Observe that, in spite of the fact that circular orbits have zero acceleration, this temperature is always \emph{larger} than the temperature measured by static detectors at the same distance from the hole (compare with \eqref{static}), and actually diverges on the photon sphere $(r=3r_{s}/2)$. These results are illustrated in fig. \ref{circfig}.

%we plot $T_{-}(v,\delta=2)$ and the corresponding adiabaticity parameter $\eta(v,\delta=2)$, to show the transient regime on a given orbit, here the last stable orbit $(r=3r_{s},\delta=2)$, and $T^{\textrm{circ}}_{-}(\delta)$ to show the divergence of the temperature as the photon sphere is approached. 

\subsection{Radial trajectories}

Radially infalling trajectories ($L=0$, $\dot{r}<0$, $\dot{v}>0$) are parametrized by their energy $E$, with $E\geq1$ corresponding to unbound states and $E<1$ to bound states. (The limiting case $E=1$ describes a detector dropped from infinity with zero velocity into the black hole.) In addition to the time-dependence of the Hawking phenomenon itself (the transient represented by $v$-dependent terms in \eqref{formulacirc}), these trajectories are intrinsically non-stationary: as the black hole is approached ($\delta\rightarrow0$) the local Riemann curvature becomes larger and larger. To avoid mixing up these two effects, from now on we will assume that the trajectory is well inside the Hawking region. This allows us to focus on the time-dependent effects arising from the trajectory itself.

%One important aspect of these infallng trajectories is that they become more and more \emph{non-stationary} as they approach the horizon. To get an understanding of the physical phenomena taking place at horizon-crossing, it is therefore important to estimate both the quasi-temperatures $T_{\pm}$ \emph{and} the corresponding adiabaticity parameters $\eta_{\pm}$ there.\footnote{We recall however that in the large frequency limit, the UDW transition rate $\dot{\R}_{\pm}(\Omega)$ will always appear thermal at temperature $T_{\pm}$, i.e. decay as $\dot{\R}_{\pm}(\Omega)\sim\Omega^{-1}e^{-\Omega/T_{\pm}}$.} We will compute the latter in the last paragraph of this section. 

\paragraph*{Outgoing modes.}

Consider first the behavior of outgoing modes. Using \eqref{va}, we have
%\begin{widetext}
\begin{eqnarray}\label{T-radial}
\f{\ddot{v}_{-}}{\dot{v}_{-}}&&\simeq\f{(\delta e^{\delta-\kappa v})^{\cdot\cdot}}{(\delta e^{\delta-\kappa v})^{\cdot}}\\&&=\f{\ddot{\delta}(1+\delta)+(\dot{\delta}-\kappa\dot{v})\big(2\dot{\delta}+\delta(\dot{\delta}-\kappa\dot{v})\big)-\kappa\delta\ddot{v}}{\dot{\delta}+\delta(\dot{\delta}-\kappa\dot{v})}\nonumber.
\end{eqnarray}
%\end{widetext}
Let us now focus on the aforementioned limiting cases: the asymptotic limit $(\delta\gg1)$ and at horizon-crossing $(\delta=0)$.

%and close to the singularity $(\delta\rightarrow-1)$.

In the asymptotic limit (which requires $E\geq1$ to exist), the system \eqref{system} gives $\dot{\delta}\sim-(E^{2}-1)^{1/2}/r_{s}$, $\dot{v}\sim E-(E^{2}-1)^{1/2}$, $\ddot{\delta}\simeq0$, and $\ddot{v}\simeq0$, hence from \eqref{T-radial} with $\delta\gg1$,
\be
\f{\ddot{v}_{-}}{\dot{v}_{-}}\Big|_{\textrm{asymp}}\simeq\dot{\delta}-\kappa\dot{v}=-\kappa\Big(E+\sqrt{E^{2}-1}\Big).
\ee
Thus, in this limit we get 
\be
T_{-}^{\textrm{asymp}}(E)=T_{H}\Big(E+\sqrt{E^{2}-1}\Big).
\ee
This formula is consistent with Hawking's original prediction, with a Doppler factor accounting for the relative motion between the detector and the black hole when $E>1$. In this regime the adiabaticity parameter $\eta_{-}^{\textrm{asymp}}(E)$ is of course vanishingly small. 

At horizon-crossing $(\delta=0)$, on the other hand, the equations of motion \eqref{system} give $\dot{\delta}=-E/r_{s}$ and $\dot{v}=1/2E$, and, by differentiation of the radial equation, $\ddot{\delta}=-1/2r_{s}^{2}$. Plugging this into \eqref{T-radial} now gives 
\be
\f{\ddot{v}_{-}}{\dot{v}_{-}}\Big|_{\textrm{hor}}=\f{\ddot{\delta}+2\dot{\delta}(\dot{\delta}-\kappa\dot{v})}{\dot{\delta}}=-\f{2E}{r_{s}}=-4\kappa E
\ee
and therefore 
\be\label{resultradial1}
T_{-}^{\textrm{hor}}(E)=4ET_{H}.
\ee
Thus, not only is the quasi-temperature not zero on the horizon, but for unbound states $(E\geq1)$ it is actually \emph{larger} than the Hawking temperature perceived by static observers at infinity. The formula \eqref{resultradial1} is consistent with the result of \cite{Barbado2011}.\footnote{Considering an ``observer freely falling from infinity'', they find that ``in the last stages of his approach to the horizon, and surprisingly at first sight, the effective temperature rises reaching exactly four times HawkingÕs temperature'' \cite{Barbado2011}.} 

We have stressed that the near-horizon regime is not stationary, hence that whether \eqref{resultradial1} can be interpreted as a temperature depends on the value of $\eta_{-}(E)$ there. Further differentiation of \eqref{system} and \eqref{T-radial} gives
\be
\eta_{-}^{\textrm{hor}}(E)=\f{\pi}{4}\left(2+\f{1}{E^{2}}\right).
\ee
This number is of order $1$ for all unbound trajectories $(E\geq1)$, and never smaller than $\pi/2\simeq1.6$: detectors with frequency $\Omega\simeq T_{-}^{\textrm{hor}}(E)$ will \emph{not} confuse the vacuum state with a thermal state. Large-frequency detectors ($\Omega\gg T_{-}^{\textrm{hor}}(E)$), on the other hand, will not be able to make this difference.

\paragraph*{Incoming modes.}

Let us now consider the behavior of \emph{incoming} modes. As we saw earlier, these do not couple to UDW detectors on stationary trajectories (static or circular); it may therefore seem appropriate to assume that the same holds along radial trajectories. This is indeed the case in the asymptotic limit, where we have seen that $\ddot{v}\simeq0$---but what about the near-horizon regime?

We have obtained $\dot{v}=1/2E$ on the horizon. To get the corresponding value of $\ddot{v}$, we differentiate the second equation of \eqref{system}. On the horizon, this gives $\ddot{v}=-\kappa/4E^{2}$, and therefore 
\be\label{Tminushor}
T_{+}^{\textrm{hor}}(E)=\f{T_{H}}{2E}.
\ee
For unbound states $(E\geq1)$, this quasi-temperature is always smaller than that of outgoing modes. The situation is different for bound states $(E<1)$: for any $E<1/\sqrt{8}$ we have $T_{+}^{\textrm{hor}}(E)>T_{-}^{\textrm{hor}}(E)$, i.e. the incoming modes \emph{dominate} the outgoing modes, and in the limit where $E\rightarrow0$ (detectors dropped with zero velocity from close to the horizon), the quasi-temperature of incoming modes grows \emph{larger and larger}. 

To assess the meaning of this result, we must (as before) consider the value of the adiabaticity parameter at horizon-crossing. We get
\be
\eta_{+}^{\textrm{hor}}(E)=2\pi\vert1-8E^{2}\vert.
\ee
Thus, just like for outgoing modes, $\eta_{+}^{\textrm{hor}}$ is bounded from below by a number of order one (actually $2\pi\simeq6.2$): that is, the adiabatic approximation is \emph{not} good in this regime, and \eqref{Tminushor} can only be interpreted as the ultraviolet decay rate of detector spectra. This notwithstanding, the fact that $T_{+}^{\textrm{hor}}(E)$ becomes large as $E\rightarrow0$ indicates that, in this regime, ingoing modes play a key role in the Hawking process. 

%We will see in the next section that this irreducible non-adiabaticity at the Schwarzchild horizon is due to the presence of curvature.  

\begin{figure}[t!]
\begin{center}
\includegraphics[scale=1.1]{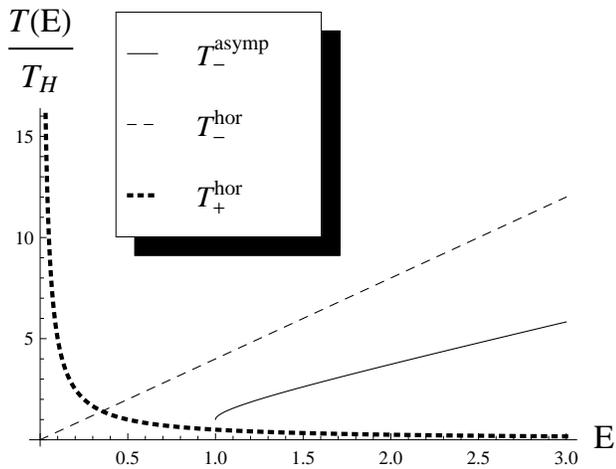}
\caption{Radially infalling trajectories: quasi-temperature of outgoing and incoming modes in different regimes, as functions of the energy $E$. For small energies $E\ll1$ (bound states), incoming modes dominate.}
\label{energy}
\end{center}
\end{figure}

%\paragraph*{Adiabaticity parameters.} We close this section on radially infalling orbits by computing the values of the adiabaticity parameters $\eta_{\pm}(E)$ at horizon-crossing. These are obtained by differentiating \eqref{T-radial} once and the equations of motion \eqref{system} twice, and evaluating both at $\delta=0$. 

%
%\begin{figure}[t!]
%\centering
%\hspace{-1cm}
%\includegraphics[scale=1]{figang}
%\caption{Effect of angular momentum: quasi-temperature of incoming modes at horizon-crossing as functions of $E$ and $L$.}
%\label{figang}
%\end{figure}

\subsection{Inspiral orbits}

It is not difficult to generalize the above results to inspiral trajectories with non-zero angular momentum. As before, the equations of motion \eqref{system} gives $\dot{\delta}$, $\ddot{\delta}$, etc., as functions of $E$ and $L$ in the various regimes, and using \eqref{T-radial} it is a simpler matter to derive $T_{\pm}^{\textrm{hor}}(E,L)$. We do not repeat this computation here; the results are 
\be
T_{-}^{\textrm{hor}}(E,L)=4ET_{H}
\ee
and 
\be
T_{+}^{\textrm{hor}}(E,L)=\f{T_{H}}{2E}\left|1+\f{L^{2}}{r_{s}^{2}}\Big(1-\f{8E^{2}}{1+L^{2}/r_{s}^{2}}\Big)\right|
\ee
and similar formulas for $\eta^{\textrm{hor}}_{\pm}$ which we do not give here. Thus, while the quasi-temperature of outgoing modes at horizon crossing is independent of the angular momentum of the trajectory, the quasi-temperature of incoming modes is not. Instead, it is given by the above (non-monotonous) function of $E$ and $L$.%, which we plot in fig. \ref{figang}.

\section{Hawking flux along radial trajectories}\label{flux}

It is well-known that the response of UDW detectors to vacuum radiation is \emph{a priori} independent from the existence of a non-zero energy-momentum tensor $\langle T_{ab}\rangle$.\footnote{In the Unruh effect, for instance, one has $\mathcal{R}\neq0$ but $\langle T_{ab}\rangle=0$.} In the Hawking effect, however, it turns out that $\langle T_{ab}\rangle\neq0$ after the collapse: at future infinity, one finds $\langle T_{uu}\rangle=\pi T_{H}^{2}/12$ (and all other components zero), showing that a Schwarzschild black hole indeed produces a stationary outgoing flux---that it actually \emph{evaporates}.   

What flux would infalling observers measure at horizon-crossing? Given the intriguing results we have obtained for the response function of UDW detectors in the limit $E\rightarrow0$, it is interesting to consider also the flux
\be
\mathcal{F}(E)=-\langle T_{ab}\rangle\, u^{a}n^{b}
\ee
along radially infalling trajectories $u^{a}(E)$ as a function of their energy $E$, say in the direction $n^{b}$ orthogonal to $u^{a}$ in the outwards direction. This quantity has been considered previously by several authors, see \cite{Ford:1993wj,Paranjape:2013vi,Anonymous:rYzr9qbR}, and has two advantages over UDW response functions:
\begin{itemize}
\item
It is \emph{local}: one does not require a trajectory extending sufficiently far in the past of the detection event to define $\mathcal{F}(E)$.
\item
It is \emph{exact}: one does not need to rely on an intricate approximation scheme (such as the adiabatic expansion of UDW response functions) to compute $\mathcal{F}(E)$. 
\end{itemize}
Is $\mathcal{F}(E)$ also divergent as $E\rightarrow0$?

To answer this question, we can indeed rely on standard results on vacuum energy-momentum tensor in two-dimensional spacetimes: given null coordinates $(v_{+},v_{-})$, the vacuum energy-momentum tensor can be written as \cite{Davies:1976uk,Davies:1977td, Birrell1982}
\begin{eqnarray}\label{formulaflux}
\langle T_{v_{\pm}v_{\pm}}\rangle&=&-\f{1}{12\pi}C^{1/2}\pp_{v_{\pm}}^{2}C^{-1/2}\nonumber\\
\langle T_{v_{+}v_{-}} \rangle&=&\langle T_{v_{-}v_{+}} \rangle=\f{RC}{96\pi},
\end{eqnarray}
where $C$ is the conformal factor such that 
\be
ds^{2}=-C(v_{+},v_{-})dv_{+}dv_{-}
\ee 
and $R=-4C^{-1}\pp_{v_{+}}\pp_{v_{-}}\ln C$ the two-dimensional scalar curvature.% in the $(v,r)$ sector of spacetime (which is not zero in Schwarzschild spacetime).
\begin{figure}[t!]
\centering
%\hspace{-1cm}
\includegraphics[scale=1.11]{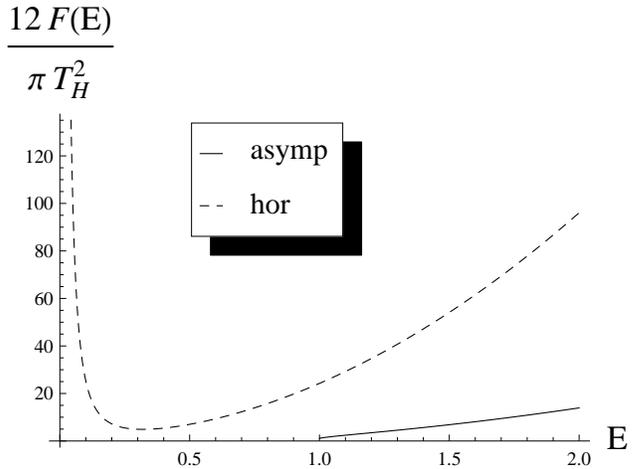}
\caption{The flux $\mathcal{F}(E)$ perceived by radially infalling observers with energy $E$, both in the asymptotic region (``asym'') and at horizon-crossing (``hor''). Compare with Fig. \ref{energy}.}
\label{fluxfig}
\end{figure}

In our case---which, by virtue of our assumption of spherical symmetry, is effectively two-dimensional---the factor $C(v_{+},v_{-})$ can be obtained from the Vaidya metric \eqref{vaidya} by inverting the relation \eqref{va} between $v_{-}$ and $(v,r)$. This gives
\be
C(v_{+},v_{-})=\f{a-1}{a}\f{W\big(-ae^{-a+\kappa v_{+}}\big)}{1+W\big(-ae^{-a+\kappa v_{+}}\big)}
\ee
where $a\equiv1+\kappa v_{-}$. Applying the formulas \eqref{formulaflux} and considering the $v_{-}\rightarrow-1/\kappa$ limit (Hawking region) for $\kappa v\gg1$ (late times), we compute
\begin{eqnarray}\label{tensor}
\langle T_{v_{+}v_{+}}\rangle^{\textrm{hor}}&=&-\f{\pi T_{H}^{2}}{12}\nonumber\\
\langle T_{v_{-}v_{-}}\rangle^{\textrm{hor}}&\sim&\f{\pi T_{H}^{2}}{2}e^{2\kappa v}.
\end{eqnarray}
Note that the incoming component $\langle T_{v_{+}v_{+}}\rangle$ at the horizon is minus the outgoing component $\langle T_{v_{-}v_{-}}\rangle$ at infinity. 

Contracting \eqref{tensor} with the radial $4$-velocity $u^{a}$ (with components $\dot{v}=1/2E$ and $\dot{v}_{-}=2Ee^{-\kappa v}$ at the horizon) and its outwards-pointing unit normal $n^{b}$ (with components $n_{+}=1/2E$ and $n_{-}=-2Ee^{-\kappa v}$), we arrive at
\be\label{fluxresult}
\mathcal{F}^{\textrm{hor}}(E)=\pi T_{H}^{2}\Big(2E^{2}+\f{1}{48E^2}\Big). 
\ee
This describes an outgoing flux, which---like the quasi-temperature $T_{-}^{\textrm{hor}}$---diverges when $E\rightarrow0$. This expression is to be compared with the asymptotic flux, given by 
\be
\mathcal{F}^{\textrm{asymp}}(E)=\f{\pi T_{H}^{2}}{12}\left(E+\sqrt{E^{2}-1}\right)^{2}. 
\ee
We plot both functions of $E$ in Fig. \ref{fluxfig}. Numerically, an observer dropped with zero velocity from e.g. $r=100r_{s}/99$ (i.e. such that $E=0.1$), the flux given by \eqref{fluxresult} is $\simeq25$ times larger than the Hawking value $\pi T_{H}^{2}/12$. 

\section{Role of curvature in near-horizon Hawking radiation}\label{curv}

We have found that UDW detectors crossing the horizon of a Schwarzschild black hole will in general record non-zero quasi-temperatures and flux. How much of this effect is due to the structure of the collapse vacuum, and how much to the mere fact that spacetime is curved on the horizon? It is impossible to answer this question within the framework of Schwarzschild black holes, because both their surface gravity (hence $T_{H}=\kappa/2\pi$) and their Riemann curvature on the horizon are controlled by the same parameter, namely the mass. To get some insight into the role played by curvature in the Hawking effect, we therefore consider artificial black holes with non-Schwarzschild geometry, in the spirit of the ``analogue gravity'' program \cite{Barcelo2011}.

\subsection{Artificial black holes with flat horizons}

Since we are interested in models of gravitational collapse, we shall consider generalizations of the Vaidya metric of the form
\be
ds_{f}^{2}=-\Big(\Theta(-v)+\Theta(v)f(r)\Big)dv^{2}+2dvdr+r^{2}d\Omega^{2}
\ee
where $f(r)$ could in principle be essentially any function of $r$. In particular, a function $f(r)$ such that $(i)$ $f(r_{s})=1$, $(ii)$ $f'(r_{s})=2\kappa$, $(iii)$ $\lim_{r\rightarrow\infty}f(r)=1$ and $(iv)$ $\lim_{r\rightarrow0}f(r)=-\infty$ will yield a black with the same asymptotic structure and surface gravity $\kappa$ as the Schwarzschild black hole, but with a different Riemann curvature distribution. Here we will focus on an artificial collapse metric of the form $f(r)=(r/r_{s}-1)\Psi(r)$, where $\Psi(r)$ is constant near the horizon but ensures that conditions $(iii,iv)$ above is satisfied. This corresponds to a spacetime where curvature (in the $(v,r)$ sector) is concentrated on two locations: at the shell ($v=0$), where it has a $\delta$-function singularity, and at large radii (say $r\geq r_{c}$), where the derivatives of $\Psi(r)$ start taking non-zero values. So long as we only consider trajectories contained within the domain $\mathcal{F}\equiv\{v>0,\ r<r_{c}\}$, we can think of this model as describing a ``flat black hole''. Does this spacetime radiate in the same way as the Schwarzschild black hole?

Repeating the computation leading to \eqref{va} in the Vaidya case, we find for this ``flat black hole''
\be\label{fl}
v_{-}(v,r)=-2r_{s}(1+\delta\,e^{-\kappa v}).
\ee
Just like the Vaidya case, this equation defines a ``Hawking region'' where $v_{-}\simeq-\kappa^{-1}$; the only difference with the Schwarzschild case is that this region grows relatively faster as a function of $v$, as can be confirmed by comparing the conditions $\delta e^{-\kappa v}\ll1$ (flat) and $\delta e^{\delta-\kappa v}\ll1$ (Vaidya).

\subsection{Quasi-temperatures at the flat horizon}

To estimate the quasi-temperatures measured by UDW detectors in geodesic motion around this flat black hole, we repeat the steps in sec. \ref{geo}: write the geodesic equation for trajectories with orbital parameters $(E,L)$, which is now
\be\label{systemflat}
\left
\{
\begin{array}{c}
\dot{r}^{2}+f(r)(1+L^{2}/r^{2})=E^{2}
\\
-f(r)\dot{v}^{2}+2\dot{v}\dot{r}+L^{2}/r^{2}=-1,
\end{array}
\right.
\ee
and from \eqref{fl} compute
\be\label{T-flat}
\f{\ddot{v}_{-}}{\dot{v}_{-}}=\f{\ddot{\delta}-2\kappa\dot{v}\dot{\delta}-\kappa\delta(\ddot{v}+\kappa\dot{v}^{2})}{\dot{\delta}-\kappa\delta\dot{v}}.
\ee
Clearly, in the asymptotic limit $r\rightarrow\infty$, these equations lead to the same results as in the Schwarzschild case. This confirms that asymptotic Hawking radiation is a global phenomenon which does not depend on the actual curvature distribution around the black hole. 

Consider however the case of UDW detectors in the near-horizon region, where $f(r)\simeq r/r_{s}-1=\delta$. In this regime, \eqref{systemflat} gives the same values for $\dot{\delta}$, $\dot{v}$ and $\ddot{\delta}$ as in the Schwarzschild case, and in particular $\ddot{\delta}=2\kappa\dot{v}\dot{\delta}$, but now $\ddot{v}=-\kappa\dot{v}^{2}$. Thus, all terms on the numerator of \eqref{T-flat} cancel, and therefore 
\be
T^{\textrm{near-hor}}_{-}(E,L)=0.
\ee 
The superscript indicates that this identity holds not just on the horizon, but in the whole near-horizon region where the black hole is flat. In this regime, the outgoing modes do \emph{not} couple to UDW detectors. This shows that, for these to become thermal, there must be a region where the Riemann curvature is non-zero between the horizon and the observer (even if its actual distribution in space is irrelevant). 

As for the incoming modes, the relationship $\ddot{v}=-\kappa\dot{v}^{2}$ gives
\be
T^{\textrm{hor}}_{+}(E,L)=\f{T_{H}}{2E}\f{(L^{2}+r_{s}^{2})}{r_{s}^{2}}.
\ee
The incoming modes, which unlike the outgoing modes must have encountered some curvature on their way to the horizon, do couple to UDW detectors, even more so that $L$ is large and $E$ small. 

We have seen at the end of the previous section that there exists an ``irreducible non-adiabaticity'' of horizon-crossing radial geodesics in the Schwarzschild spacetime. This ``irreducible non-adiabaticity'' is also present in the flat case, as can be confirmed by the computation of $\eta_{+}^{\textrm{hor}}$: sticking for simplicity to $L=0$, we get
\be
\eta_{+}^{\textrm{hor}}(E)=2\pi. 
\ee
This irreducible non-adiabaticity is thus independent of the local geometry of the horizon; it is an intrinsic feature of Hawking radiation as perceived by freely-falling near-horizon observers. 

\subsection{Flux at the horizon}

We close this section by repeating the computation of the flux measured by infalling observers $\mathcal{F}(E)=-\langle T_{ab}\rangle u^{a}n^{b}$. In the flat case, the relation \eqref{fl} gives $C(v_{+},v_{-})=e^{\kappa v}$, hence from the Davies-Fulling-Unruh formula \eqref{formulaflux}
\begin{eqnarray}\label{tensorflat}
\langle T_{v_{+}v_{+}}\rangle=-\f{\pi T_{H}^{2}}{12};\ 
\langle T_{v_{-}v_{-}}\rangle=\langle T_{v_{-}v_{+}}\rangle=0.
\end{eqnarray}
and therefore
\be
\mathcal{F}^{\textrm{hor}}(E)=\f{\pi T_{H}^{2}}{48E^2}. 
\ee
Again, we find that flattening the horizon cancels the effect of outgoing modes at the horizon, but not that of ingoing modes. The result is an outgoing flux which diverges as $E\rightarrow0$.

\section{Discussion and conclusion}\label{disc}

In this paper, we have considered Hawking radiation from the perspective of UDW detectors evolving on non-asymptotic trajectories. When the trajectory is not stationary, or just after the collapse, their response is not exactly thermal; by using a ``quasi-temperature'' formalism based on a suitable adiabatic expansion, we have shown how to get a basic understanding of their response, especially in the large frequency limit. This quasi-temperature formalism is very convenient from a computational perspective: formula \eqref{quasi} is explicit and can be straightforwardly applied to any trajectory of interest.  

Following this approach, we have obtained several interesting results, which go beyond the lore that Hawking radiation simply reduces to Unruh radiation at the horizon, or that ``the equivalence principle is restored'' \cite{Singleton:2011vj} there (that is, assuming that the Hawking effect violates it in some sense\footnote{We probably would not have described the Hawking effect in those terms anyhow.}). They are:
\begin{itemize}
\item
In spite of the fact that these trajectories are not accelerated, the temperature perceived on circular orbits is always higher than that on static trajectories at the same distance from the hole, and diverges on the photon sphere.
\item
In the near-horizon region, a freely-falling detector couples to both the outgoing \emph{and} the incoming modes of the field, to which it associates two different quasi-temperatures,\footnote{At least in the $s$-wave approximation discussed in this paper.} depending on its energy and angular momentum. Both can be arbitrarily high (though in different regimes: for fast and slow moving detectors respectively). This singular behavior of Hawking radiation in the $E\rightarrow0$ limit\footnote{This can be thought of as a divergent Doppler effect: the Doppler factor between two geodesic observers with energies $E$ and $E'$ diverges like $\ln(E'/E)$ when $E\rightarrow0$. We thank Luis Garay and Jorma Louko for this observation.}, is confirmed by a flux computation. 
\item
While the thermality of outgoing modes relative to near-horizon geodesic observers can be traced back to the spacetime curvature, this is not the case for incoming modes: these appear thermal (both in terms of UDW response functions and of fluxes) to infalling observers also in spacetimes with flat horizons. In both cases, there exists an irreducible non-adiabaticity at the horizon, in the sense that $\vert\dot{T}_{\pm}\vert\sim T_{\pm}^{2}$ at the horizon. 
\end{itemize}
In a nutshell: far from a no-particle vacuum, a detector dropped with zero velocity from near the horizon will in fact record intense Hawking radiation in the ingoing sector. 

%Far 
%how a radiating black hole is \emph{not} like a black body---not just because of grey body factors, but also in other, more fundamental ways.

\acknowledgments

SS is supported by an SPM grant from the Council of Scientific and Industrial Research (CSIR), India. We are grateful to Thanu Padmanabhan for suggesting to study Hawking radiation from the perspective of infalling detectors and for kindly hosting one of us (MS) at IUCAA, India, during a research visit. We also thank Antonin Coutant, Ahmed Youssef and Carlo Rovelli for useful conversations. 
\appendix

\section{Derivation of \eqref{thermalspectrum}}
\label{thermalresult}
From Eq.~\eqref{vminusspectrum} and the fact that $v_{+}(\tau)$ is linear in $\tau$ in the Hawking region, the Wightman function in \eqref{2d} can be expressed as
\be
G(\tau,\tau - s) \propto  \ln(s) + \ln\left(e^{-\f{\kappa}{2}(2\tau -s)}\right) +\ln\left( \sinh\left(\f{\kappa}{2}s\right)\right)
\ee
When substituted in \eqref{response1}, the first and the second term on integration by parts contribute $\Theta(-\Omega)$ or Dirac delta in $\Omega$ and its derivatives which have support on $\Omega = 0$. Since we are looking at cases for finite energy level separation between two levels, that is, $\Omega > 0$, these contributions (in fact any polynomial in $s$) are irrelevant. As for the the third term, we observe that it is periodic in $s$ with an imaginary period $2i\pi/\kappa$; it contribution to \eqref{response1} can be computed explicitly by residue calculus, by using
\be
\sinh\left(x\right) = x \prod^\infty_{m=1}\f{(m\pi - ix)(m\pi + i x)}{m^2\pi^2}
\ee
and summing over the poles. This results in \eqref{thermalspectrum}.

%and
%\be
%\int_\sigma^\infty d\omega \f{e^{-i\omega x}}{\omega\left(e^{\beta\omega} -1\right)} \xrightarrow{\sigma \rightarrow 0} -\ln\left[\prod^\infty_{m=1}\sigma e^\gamma (\beta m + i x)\right] 
%\ee
%where $\gamma$ is the Euler constant, we can write the relevant part as
%\begin{align}
%\int_0^\infty ds\, e^{-i\Omega s}&\ln\left( \sinh\left(\f{\kappa}{2}s\right)\right) \nonumber\\
%&\sim 2\int_0^\infty ds\, e^{-i\Omega s}\int_0^\infty d\omega \f{\cos\left(\omega\kappa s/2\right)}{\omega(e^{\pi\omega} -1)}.
%\end{align}
%Expressing the cosine as sum of exponentials, interchanging the order of integration to integrate over $\omega$ gives two Dirac deltas. Keeping only the positive contribution to $\Omega$ gives the result
%\be
%\f{\kappa}{2\Omega\left(e^{2\pi\Omega/\kappa} -1\right)}.
%\ee
\section{Adiabatic expansion of UDW-like response functions}

%\section{Schwarzschild radial metric in $(v_{+},v_{-})$ coordinates}

The validity of the adiabatic approximation mentioned in sec. \ref{quasi} has been studied in some detail \cite{Obadia:2007ds,Kothawala:2009aj,Barcelo:2010xk} in the context of non-uniformly accelerated UDW detectors in flat spacetime (approximate Unruh effect). In this setup, Barbado and Visser \cite{Barcelo:2010xk} have devised an ``adiabatic expansion'' for the UDW transition rates in terms of the quantities $a^{(n)}(\tau)/a(\tau)^{n+1}$, where $a(\tau)$ is the instantaneous acceleration of the detector. 

In this appendix, we describe the adiabatic expansion of any functional of a slowly varying function $k(\tau)$ of the form
\be\label{functional}
R(\tau,\Omega;k]\equiv\int_{-\infty}^{\infty}ds\,K(\Omega s)\,g(\Delta_{\tau,s}[k]).
\ee
where $K$, $g$ are any functions such that the above integral is well-defined, and $\Delta_{\tau,s}[k]$ is the functional 
\be
\Delta_{\tau,s}[k]\equiv \f{v(\tau)-v(\tau-s)}{\dot{v}(\tau)s},
\ee 
with $v(\tau)$ any solution of the differential equation 
\be\label{ODE}
\ddot{v}(\tau)=-k(\tau)\dot{v}(\tau). 
\ee
The UDW response functions of the kind discussed in this paper obviously fit this scheme, with $K(z)=\exp(-iz)$, $g(z)=\ln(z)$, and $k$ (divided by $2\pi$) the quasi-temperature function.

The functional $R(\tau,\Omega;k]$ defined by \eqref{functional} may be seen as a function of $\tau$, $\Omega$, and the infinitely many derivatives $k^{(n)}(\tau)$ of $k$ at a given instant $\tau$, or---better---as $\Omega^{-1}$ times the dimensionless function 
\be\label{dimensionless}
R(\nu_{0},\nu_{1},\dots)\equiv\int_{-\infty}^{\infty}du\,K(u)\,g(\Delta_{\tau,u/\Omega}[k])
\ee 
of the dimensionless ratios $\nu_{n}\equiv k^{(n)}(\tau)/\Omega^{n+1}$. If we denote $R_{0}(\nu_{0})$ the value of $R(\nu_{0},\nu_{1},\dots)$ in the particular case where $k(\tau)=k$ is constant, the question we wish to answer is: in the general case where $k(\tau)$ is not constant, can $R(\nu_{0},\nu_{1},\dots)$ be systematically expanded about $R^{*}(\nu_{0})$, with each corrective term depending on only finitely many $\nu_{n}$'s? 

The key idea to get started with this question, which we found in \cite{Barcelo:2010xk}, consists in introducing a scaling parameter $\alpha$ and $k^{(\alpha)}(\tau-s)\equiv k(\tau-\alpha s)$. Denoting $\nu^{(\alpha)}_{n}\equiv\alpha^{n}\nu_{n}$ the corresponding scaled dimensionless ratios, we have
\be\label{taylor}
R(\nu_{0},\nu_{1},\dots)=\sum_{m=0}^{\infty}\f{1}{m!}\f{d^{m}}{d\alpha^{m}}R(\nu^{(\alpha)}_{0},\nu^{(\alpha)}_{1},\dots)\big|_{\alpha=0}.
\ee
Using Fa\`a di Bruno's formula (chain rule for higher derivatives), we may write the $m$-th term in this expansion as 
\be\label{faa}
\f{1}{m!}\sum_{j=0}^{m}\int_{-\infty}^{\infty}du\,K(u)g^{(j)}(\Delta_{0})B_{m,j}(\Delta_{1},\cdots,\Delta_{m-j+1}).
\ee
where $B_{n,j}$ are the Bell polynomials\footnote{The Bell polynomials are defined by $B_{m,k}(z_{1},\cdots,z_{m-k+1})=m!\sum\prod_{i=1}^{m-k+1}\f{1}{j_{i}!}(\f{z_{i}}{i!})^{j_{i}}$, where the sum runs over all sequences of non-negative integers $(j_{i},\cdots,j_{m-k+1})$ such that $j_{1}+\cdots+j_{m-k+1}=k$ and $j_{1}+2j_{2}+\cdots+(m-k+1)j_{m-k+1}=m$.}, and 
\be\label{Dd}
\Delta_{j}\equiv\partial_{\alpha}^{j}\Delta_{\tau,u/\Omega}[k^{(\alpha)}]\big|_{\alpha=0}.
\ee
In the next paragraph, we will show that $\Delta_{j}$ is given by 
\be\label{D}
\Delta_{j}=\f{\nu_{j}}{\nu_{0}^{j+1}}\delta_{j}(\nu_{0}u)
\ee 
for some function $\delta_{j}$. This result calls for several comments:
\begin{itemize}
\item
The $m$-th term in the Taylor expansion \eqref{taylor}, given by \eqref{faa}, is is a function $R_{m}(\nu_{0},\cdots,\nu_{m})$ of the first $m+1$ parameters $\nu_{n}$ only, viz.
\be\label{adiab}
R(\nu_{0},\nu_{1},\dots)=\sum_{m=0}^{\infty}R_{m}(\nu_{0},\cdots,\nu_{m}).
\ee

\item
The relevant parameters controlling the convergence of the adiabatic expansion \eqref{adiab} are indeed $\nu_{j}/\nu_{0}^{j+1}=k^{(j)}(\tau)/k(\tau)^{j+1}$, as observed in \cite{Barcelo:2010xk} in the context of the Unruh effect; in particular, we have for each $m\geq1$
\be
R_{m}(\nu_{0},\cdots,\nu_{m})=\mathcal{O}\Big(\sup_{1\leq j\leq m}\f{\nu_{j}}{\nu_{0}^{j+1}}\Big).
\ee
When $k(\tau)$ is a polynomial function of $\tau$ (so that only finitely many $\nu_{n}$'s are non-vanishing), we are assured that
\be
R(\nu_{0},\nu_{1},\cdots)=R_{0}(\nu_{0})+\mathcal{O}\Big(\sup_{j\geq1}\f{\nu_{j}}{\nu_{0}^{j+1}}\Big).
\ee
In other words, the adiabatic approximation is good when all the expansion parameters $\nu_{j}/\nu_{0}^{j+1}$ are small.
\item
The adiabatic expansion is not a high-frequency expansion; indeed, the functions $\delta_{j}(\nu_{0}u)=\delta_{j}(k(\tau)u/\Omega)$ can well be non-polynomial in $1/\Omega$.
\end{itemize}

To prove \eqref{D}, we begin by expressing the argument of $g$ in \eqref{dimensionless} explicitly in terms of the dimensionless ratios $\nu_{n}$'s. From \eqref{ODE}, we have
\begin{eqnarray}
\Delta_{\tau,u/\Omega}[k]=\f{\Omega}{u}\int_{-u/\Omega}^{0} d\sigma\,\exp\int_{\sigma}^{0}d\rho\,k(\tau+\rho).
\end{eqnarray}
Expanding $k(\tau+\rho)$ in powers of $\rho$ and making the change of variable $\mu=\Omega\sigma$ in the $\sigma$-integral, this gives
\be\label{deltaj}
\Delta_{\tau,u/\Omega}[k]=\f{1}{u}\int_{0}^{u}d\mu\,\exp\sum_{n=0}^{\infty}\f{\mu^{n+1}\nu_{n}}{(n+1)!}.
\ee
Next, we consider the effect of scaling by $\alpha$ on \eqref{deltaj}, 
\be
\Delta_{\tau,u/\Omega}[k^{(\alpha)}]=\f{1}{u}\int_{0}^{u}d\mu\,\exp\sum_{n=0}^{\infty}\f{\mu^{n+1}\alpha^{n}\nu_{n}}{(n+1)!}
\ee
and compute its $j$-th derivative with respect to $\alpha$, as in \eqref{deltaj}. The $0$-th derivative is easily found to be
\be\label{Delta0}
\Delta_{0}=\f{e^{\nu_{0}u}-1}{\nu_{0}u}
\ee
and for, $j\geq1$,
\be
\Delta_{j}=\f{\nu_{j}}{j+1}\pp_{\nu_{0}}^{j+1}\Delta_{0}.
\ee
Computing explicitly the derivatives of \eqref{Delta0}, we arrive at \eqref{D} with
\be\label{deltaj}
\delta_{j}(z)=\f{(-1)^{j+1}j!}{z}\left(e^{z}\sum_{r=0}^{j+1}\f{(-1)^{r}}{r!}z^{r}-1\right).
\ee
%This result confirms our claim that $\Delta_{j}$ is $\nu_{j}$ times a function of $\nu_{0}$ and $u$ only. Plugging \eqref{Deltaj} in\eqref{faa} turns \eqref{adiab} into an explicit adiabatic expansion, which can be computed order by order in the $\nu_{n}'s$. 

We close this appendix by evaluating explicitly the first corrective term $R_{1}$ in the particular case where $K(z)=e^{-iz}$ and $g(z)=\ln z$, namely
\be
 R(\tau,\Omega;k]=\Omega^{-1}R(\nu_{0},\cdots)\equiv\int_{-\infty}^{\infty}ds\,e^{-i\Omega s}\,\ln(\Delta_{\tau,s}[k]).
 \ee
The real part of this integral is the UDW response function along a trajectory with quasi-temperature $T=k/2\pi$, studied in the main body of this paper. As is well-known, the $0$-th order $R_{0}(\nu_{0})$ is the standard two-dimensional thermal spectrum
\be
R_{0}(\nu_{0})=\f{2\pi}{e^{2\pi/\nu_{0}}-1}.
\ee
From \eqref{faa}, \eqref{Dd} and \eqref{deltaj}, we have 
\begin{multline}
R_{1}(\nu_{0},\nu_{1})=\f{\nu_{1}}{\nu_{0}^{2}}\times\\ \int_{-\infty}^{\infty}du\, e^{-iu}\f{e^{\nu_{0}u}(1-\nu_{0}u+(\nu_{0}u)^{2}/2)-1}{e^{\nu_{0}u}-1}
\end{multline}
The integrand has simple poles at $u_{n}=2i\pi n/\nu_{0}$ for each integer $n\neq0$. Closing the contour in the lower half-plane and summing over $n\leq-1$, we obtain 
\be
R_{1}(\nu_{0},\nu_{1})=\f{\nu_{1}}{\nu_{0}^{2}}\left(\f{\pi^{2}\Big(1+i\pi\coth(\pi/\nu_{0})\Big)}{\nu_{0}\sinh^{2}(\pi/\nu_{0})}\right).
\ee
Reinstating the dimensionful variables $k(\tau)$, $\dot{k}(\tau)$ and $\Omega$ and writing $T=k/2\pi$, this gives
\begin{eqnarray}
\textrm{Re}\{R(\tau,\Omega;k=2\pi T]\}=\f{2\pi}{\Omega(e^{\Omega/T}-1)}\Big(\,1\,+\,\nonumber\\ \f{\dot{T}(\tau)}{T(\tau)^{2}}\f{\Omega}{8\pi T(\tau)}\f{e^{\Omega/2T(\tau)}}{\sinh^{2}(\Omega/2T(\tau))}+\cdots\Big).
\end{eqnarray}
Remarkably, the second term in the brackets is \emph{exponentially} vanishing in the ultraviolet limit, showing that the adiabatic approximation is excellent in this regime, whatever the value of the parameter $\dot{T}(\tau)/T(\tau)^{2}$.  
%and hence
%\be
%R(\nu_{0},\nu_{1},\dots)=\int_{-\infty}^{\infty}du\,K(u)g\left(\f{1}{u}\int_{-u}^{0}d\mu\,\exp\sum_{n=0}^{\infty}\f{(-1)^{n+1}}{(n+1)!}\nu_{n}\right).
%\ee
%the generalized chain rule (Fa\`a di Bruno's formula) gives

%In the special case where $k(\tau)$ is constant, elementary dimensional analysis shows that $R(\tau,\Omega;k)$ is proportional to a function $R^{*}(k/\Omega)$ of $k/\Omega$ only. In the general case, the functional of $k$ defined by \eqref{functional} can be seen a function of all the derivatives $k^{(n)}(\tau)$ of $k$ at a given time $\tau$, and the same line of reasoning shows that $R(\tau,\Omega;k\,]$ may be cast as a function of the dimensionless parameters $k^{(n)}(\tau)/\Omega^{n+1}$ only. The adiabatic expansion of $R(\tau,\Omega;k\,]$ for a given $\Omega$ is thus best seen as an expansion is powers of $\eta(\Omega)\equiv\sup_{n\geq1}\{k^{(n)}(\tau)/\Omega^{n+1}\}$; the ``adiabatic limit'' corresponds to the case where this expansion converges quickly for any $\Omega$ larger than the characteristic frequency $k(\tau)$, so that the whole spectrum is well approximated by $R^{*}(k(\tau)/\Omega)$ (and not just its ultraviolet tail), and requires $\eta\equiv\eta(k(\tau))\ll1$. Let us now describe this expansion in detail.
%
%\section{The Lambert $W$-function}
%
%For the reader's convenience, we collect here some relevant facts about the Lambert $W$-function. 

\bibliographystyle{utcaps}
\providecommand{\href}[2]{#2}\begingroup\raggedright\endgroup
\end{document}